\documentclass[english,twocolumn,pra,superscriptaddress]{revtex4-1}
\usepackage[T1]{fontenc}
\usepackage[utf8]{inputenc}
\usepackage{refstyle}
\usepackage{amsmath}
\usepackage{amssymb}
\usepackage{graphicx}

\usepackage[pdftex, hidelinks]{hyperref}
\usepackage[dvipsnames]{xcolor}

\newcommand\myshade{85}
\colorlet{mylinkcolor}{NavyBlue}
\colorlet{mycitecolor}{NavyBlue}
\colorlet{myurlcolor}{NavyBlue}
\hypersetup{
	linkcolor  = mylinkcolor!\myshade!black,
	citecolor  = mycitecolor!\myshade!black,
	urlcolor   = myurlcolor!\myshade!black,
	colorlinks = true,
}

\makeatletter

\AtBeginDocument{}
\RS@ifundefined{subsecref}
  {\newref{subsec}{name = \RSsectxt}}
  {}
\RS@ifundefined{thmref}
  {\def\RSthmtxt{theorem~}\newref{thm}{name = \RSthmtxt}}
  {}
\RS@ifundefined{lemref}
  {\def\RSlemtxt{lemma~}\newref{lem}{name = \RSlemtxt}}
  {}

\usepackage{tikz}
\usepackage{hyperref}
\usepackage{bm}
\newcommand{\esw}{\mathcal{S}}
\newcommand{\sref}[2]{\hyperref[#1]{\ref*{#1}(#2)}}
\newcommand{\appx}[1]{Appendix \ref{#1}}

\makeatother

\usepackage{babel}
\begin{document}

\title{Superradiant parametric conversion of spin waves}

\author{Mateusz Mazelanik}
\email{m.mazelanik@cent.uw.edu.pl}

\selectlanguage{english}%

\affiliation{Centre for Quantum Optical Technologies, Centre of New Technologies,
University of Warsaw, Banacha 2c, 02-097 Warsaw, Poland}

\affiliation{Faculty of Physics, University of Warsaw, Pasteura 5, 02-093 Warsaw,
Poland}

\author{Adam Leszczy\'{n}ski}

\affiliation{Centre for Quantum Optical Technologies, Centre of New Technologies,
University of Warsaw, Banacha 2c, 02-097 Warsaw, Poland}

\affiliation{Faculty of Physics, University of Warsaw, Pasteura 5, 02-093 Warsaw,
Poland}

\author{Micha\l{} Lipka}

\affiliation{Centre for Quantum Optical Technologies, Centre of New Technologies,
University of Warsaw, Banacha 2c, 02-097 Warsaw, Poland}

\affiliation{Faculty of Physics, University of Warsaw, Pasteura 5, 02-093 Warsaw,
Poland}

\author{Wojciech Wasilewski}

\affiliation{Centre for Quantum Optical Technologies, Centre of New Technologies,
University of Warsaw, Banacha 2c, 02-097 Warsaw, Poland}

\author{Micha\l{} Parniak}
\email{m.parniak@cent.uw.edu.pl}

\selectlanguage{english}%

\affiliation{Centre of New Technologies, University of Warsaw, Banacha 2c, 02-097
Warsaw, Poland}

\affiliation{Niels Bohr Institute, University of Copenhagen, Blegdamsvej 17, DK-2100
Copenhagen, Denmark}
\begin{abstract}
Atomic-ensemble spin waves carrying single-photon Fock states exhibit
nonclassical many-body correlations in-between atoms. The same correlations
are inherently associated with single-photon superradiance, forming
the basis of a plethora of quantum light-matter interfaces. We devise
a scheme allowing the preparation of spatially-structured superradiant
states in the atomic two-photon cascade using spin-wave light storage.
We thus show that long-lived atomic ground-state spin waves can be
converted to photon pairs opening the way towards nonlinear optics
of spin waves via multi-wave mixing processes.
\end{abstract}
\maketitle

\section{Introduction}

Spatially extended atomic ensembles are a particularly versatile medium
allowing generation and control of light with widely varying properties. Fulfillment of the phase-matching (PM) condition 
facilitates efficient generation, storage and retrieval of single photons. Superradiance is inherently linked to the phase-matched emission and, in particular, spin waves (SW) that store information about light in the atomic coherence are superradiant Dicke states $N^{-1/2} \sum_j e^{i \mathbf{K} \mathbf{r}_j}|g_1 \ldots h_j \ldots g_N\rangle$~\cite{Dicke1954}.
In practice, the $\Lambda$ scheme of atomic levels, which forms the basis of the Duan-Lukin-Cirac-Zoller
quantum entanglement distribution protocol \citep{Duan2001}, is well-known for its capabilities to generate photon pairs \citep{Kuzmich2003,VanderWal2003} with both non-trivial
temporal \citep{Zhao2016} and spatially-multimode structure \citep{Parniak2017,Pu2017}. There, light is interfaced with a coherence between two metastable ground-state sublevels. 

An alternative ladder or diamond schemes allow generation of two-color
photon pairs, and attract much attention \citep{Gulati2015,Guo2017,Srivathsan2013a,Zhang2014}
also for single-photon storage \citep{Kaczmarek2018,Finkelstein2018}
and in Rydberg-blockaded media \citep{Lukin2001,Gaetan2009}. In those cases the associated SWs lie between a ground state and an excited atomic state and are intermediate steps in the generation of a photon pair. More complex manipulations of those SWs, such as temporal \citep{Mazelanik2019,Reim2012,Saglamyurek2018,Campbell2014,Yang2018}
and spatial \citep{Parniak2019} mutli-SW beamsplitters demonstrated for ground-state SWs, remain elusive. Such control would allow SW-based engineering of photon pair emission, possibly also extensible to deterministic quantum nonlinear optics based on Rydberg atoms \citep{Li2016}. 

The atomic-ensemble-based
schemes hold some advantages over spontaneous parametric down-conversion
processes (SPDC) in nonlinear crystals and optical parametric oscillators \citep{Kwiat1995,Neergaard-Nielsen:07,Takeno:07}, that also
rely on engineered PM, as the photons generated in atomic
ensembles are inherently narrowband and atom-resonant, making them
suitable for quantum metrology \citep{Qiu2016} and repeater-based
communication \citep{Duan2001,Sangouard2011,Niizeki2018}. With purely
atomic photon-pair source, potentially difficult engineering of cavity-based
SPDC can be avoided \citep{Wolfgramm2008,Wolfgramm2011}.

\begin{figure*}
	\includegraphics[width=.9\textwidth]{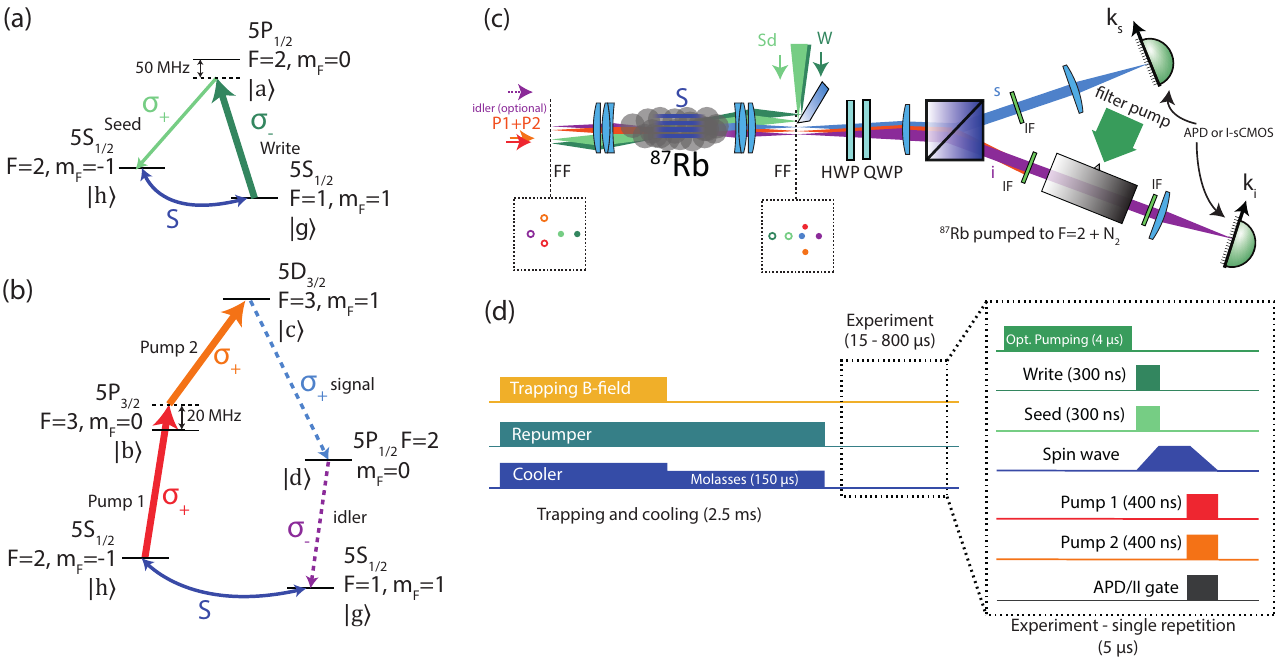}
	
	\caption{Experimental setup details. (a) Rubidium-87 levels configuration used
		to generate ground state SW via seeded Raman scattering in
		the first step of the delayed 6WM scheme. (b) Relevant energy levels
		used to generate correlated photon pairs via spontaneously induced
		two-photon decay closing the 6WM scheme. (c) Schematic of the experimental
		setup used to observe correlated photon pairs generated within the
		delayed 6WM scheme.   (d) Time sequence of the experiment detailing
		the order of all experimental steps leading to correlated photons
		pairs generation. \label{fig:exp_set}}
\end{figure*}
Here we show that the properties of a phase-matched cascaded atomic
decay can be engineered via proper preparation of the atomic SW
state. In particular, we treat the SW prepared in the Raman
process as one of the fields participating in the parametric conversion, similarly as a pump field in a spontaneous parametric down-conversion (SPDC) in nonlinear crystals \citep{Kwiat1995}. Hitherto schemes necessarily required that the atom starts
and ends the wave-mixing processes in the same ground state, as dictated
by the principles of energy and momentum conservation. This requirement
can be leveraged if the ensemble is first prepared in a superposition
state, or in other words some coherence is present.  Our scheme allows for complex engineering of two-photon emission, with atom-resonant photons without the need of cavities.

We generate a
strong atomic coherence (SW) between $|g\rangle$ and $|h\rangle$ via Raman interaction in the $\Lambda$ scheme, as depicted
in Fig. \sref{fig:exp_set}{a}. The two pump fields (P1 and P2), as
in Fig. \sref{fig:exp_set}{b}, then transfer the coherence $|g\rangle\langle h|$ via a
two-photon process to the $|g\rangle\langle c|$ involving the highest excited state. The
standard coherent two-photon decay follows by the end of which the
atom returns to its ground state $|g\rangle$. The entire process
can be viewed as a two-step spontaneous six-wave mixing (6WM) with
a time delay, previously demonstrated for classical beams \cite{Parniak2016a}, which overall conserves both momentum and energy. 

This unorthodox view leads to numerous implications, as the SW
may be manipulated in various ways, including usage of magnetic \citep{Sparkes2013,Campbell2012,Higginbottom2015},
electric \citep{Hetet2008a} and optical fields \citep{Parniak2019,Mazelanik2019,Lipka2019,Ham2017,Sparkes2010}.
Reshaping the SW structure will subsequently result in
a modified atomic state during emission of the idler photon. Such
a modification can serve to change the properties of superradiance
facilitating better understanding of this collective process, and
harnessing it in bulk atomic ensembles. From the quantum information
perspective, both polarization, temporal and spatial properties of
the generated biphoton wave-function can be modified, leading to complex
multi-degree-of-freedom entangled states. 

The ladder \citep{Lee2017,Lee2016,Jeong2017} and diamond type \citep{Parniak2015,Becerra2008,Jen2017}
systems have previously been successfully used to generate photons
with particularly good characteristics. The key advantages of these
schemes applied to Rubidium are the possibility to generate one of
the photons at telecom wavelength \citep{Chaneliere2006,Radnaev2010a,Ding2015c}
and relaxed requirements for filtering, even in room-temperature ensembles
\citep{Zhang2014,Willis2011,Park2017}, as compared with the $\Lambda$
scheme. Previous preliminary studies have suggested that excited-to-excited state transition could be used for readout of a ground-state atomic coherence \cite{Parniak2016a,Ding2012}.

 A striking feature of the two-photon decay is the superradiant
enhancement of emission \citep{Dicke1954,Srivathsan2013a,Lee2017},
observed both in ultracold and warm atomic ensembles at high optical
depths. Superradiance, while previously observed in many systems, here is shown to be controlled within the spatial domain. Our system thus contributes to recent fundamental studies
of superradiant emission in cold atoms \citep{Kuraptsev2017,Roof2016,Ortiz-Gutierrez2018}.
Among many intricacies, this behavior proves the role of atomic coherence
in the process.

\section{Experiment}

We start the  experiment by employing two coherent optical fields
(Write and Seed, both 795 nm) to generate macroscopic ground-state
atomic coherence $\rho_{hg}$ between states $|g\rangle$ and $|h\rangle$
in a cold $^{87}\mathrm{Rb}$ atomic ensemble (see Fig. \sref{fig:exp_set}{a},
and \appx{apx:esud}\ for details). This coherence, commonly called a spin wave
has a nontrivial spatial dependence $\rho_{hg}(\mathbf{r})\propto\exp(i\mathbf{K}\mathbf{r})$,
where $\mathbf{K}=\mathbf{k}_{W}-\mathbf{k}_{Sd}$ is a SW
wavevector and $\mathbf{k}_{W}$, $\mathbf{k}_{Sd}$ are wavevectors
of the Write and Seed field respectively. 

In next step we use a pair of strong laser fields (Pump 1 and Pump
2) to transform substantial part of excitations of state $|h\rangle$ into
the state $|c\rangle$, creating an excited-state SW: $\rho_{cg}(\mathbf{r})\propto\rho_{hg}\exp(i(\mathbf{k}_{P1}+\mathbf{k}_{P2})\mathbf{r})$,
where $\mathbf{k}_{P1}$, $\mathbf{k}_{P2}$ are the wavevectors of
the Pump 1 and 2 respectively. For clarity, assuming large diameters of  Pump 1\&2 beams, we define the
excited-state SW as: 
\begin{equation}
\esw(\mathbf{r})\equiv\sqrt{n(\mathbf{r})}\rho_{cg}(\mathbf{r})=\sqrt{n(\mathbf{r})}\beta e^{i\mathbf{(K+}\mathbf{k}_{P1}+\mathbf{k}_{P2})\mathbf{r}},\label{eq:esw}
\end{equation}
where the complex SW amplitude $\beta$ incorporates all
the phase shifts between pump fields, and by $n(\mathbf{r})$ we denote
the local atom density. The presence of $\esw$ enables spontaneously
induced coherent two-photon transition to the original ground state
level $|g\rangle$ closing the 6WM scheme. 

The transition happens in two steps: first the spontaneously emitted
signal photon transfers one collective excitation of $|c\rangle$ to the intermediate state $|d\rangle$, conditionally creating a Fock-state
SW between levels $|d\rangle$ and $|g\rangle$, from which
the superradiantly enhanced emission of the idler photon occurs. Thus,
the idler photon is emitted always after the signal photon.

The whole quantum amplitude $\psi(t_{s},t_{i},\mathbf{k}_{s\perp},\mathbf{k}_{i\perp})$
of the generated biphoton state exhibits two features: the spatial and temporal correlation. Here, however, we
study these features separately and thus we will define two proper
objects: the time-averaged spatial wave function $\psi_{k}(\mathbf{k}_{s\perp},\mathbf{k}_{i\perp})=\langle\psi(t_{s},t_{i},\mathbf{k}_{s\perp},\mathbf{k}_{i\perp})\rangle_{t}$,
and the temporal wave function $\psi_{t}(t_{s},t_{i})$ for fixed
pair of $\left\{ \mathbf{k}_{s\perp},\mathbf{k}_{i\perp}\right\} $. 
\begin{figure}
	\includegraphics[width=1\columnwidth]{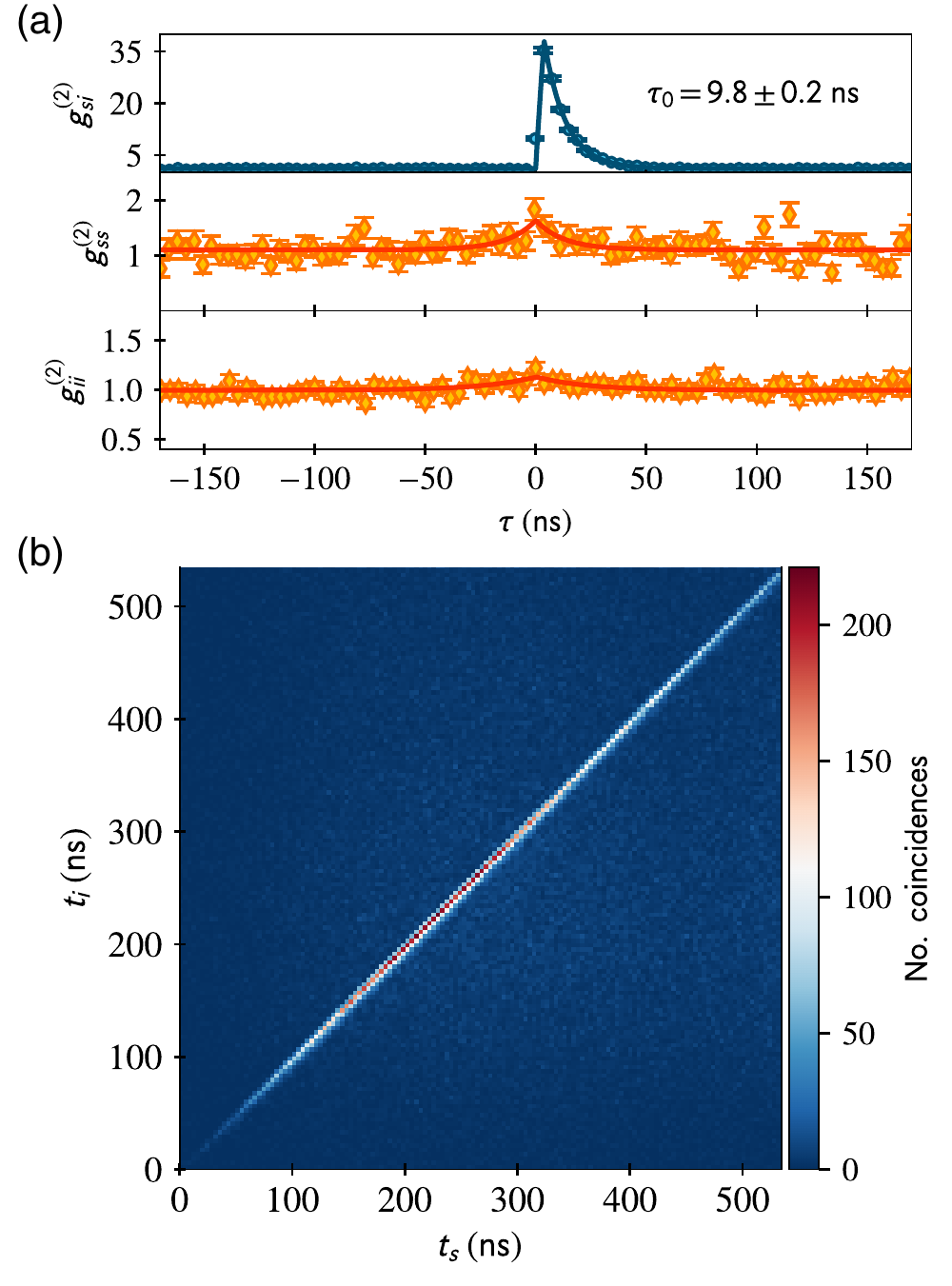}
	
	\caption{(a) Measured second order correlation functions between signal and
		idler fields $g_{si}^{(2)}(\tau=t_{i}-t_{s})$ for the two possible
		decay paths through different intermediate level $|d\rangle$ ($m_{F}=0$).
		and auto correlation functions $g_{jj}^{(2)}(\tau=t_{j1}-t_{j2})$
		for the signal ($j=s$) and idler ($j=i$) fields. Registered photon rate amounts to $6\times10^{-3}\ \mathrm{photon}/\mu\mathrm{s}$ for signal photons. (b) Raw signal--idler coincidence data for the cross-correlation accumulated during the entire pulse.\label{fig:tcorr}}
\end{figure}

From quantum perturbation theory (see \appx{apx:sd} for the derivation using Hamiltonian treatment) for weak signal
and idler fields, assuming that the atom density $n(\mathbf{r})$ is described
by the Gaussian distribution with longitudinal width equal $\sigma_{z}$
we can write the spatial part as:
\begin{equation}
\psi_{k}(\mathbf{k}_{s\perp},\mathbf{k}_{i\perp})\propto\tilde{n}(\Delta\mathbf{k}_{\perp})e^{-\Delta k_{z}^{2}\sigma_{z}^{2}},\label{eq:psi_kt}
\end{equation}
where by $\tilde{n}$ we denote the Fourier transform of the atomic
density, the transverse wavevector mismatch is defined as follows:
$\Delta\mathbf{k}_{\perp}=\mathbf{K}_{\perp}+\mathbf{k}_{P1\perp}+\mathbf{k}_{P2\perp}-\mathbf{k}_{s\perp}-\mathbf{k}_{i\perp}$,
and by 
\begin{multline}
\Delta k_{z}\equiv K_{z}+\sqrt{k_{P1}^{2}-k_{P1\perp}^{2}}+\sqrt{k_{P2}^{2}-k_{P2\perp}^{2}}\\- \sqrt{k_{s}^{2}-k_{s\perp}^{2}}-\sqrt{k_{i}^{2}-k_{i\perp}^{2}}
\end{multline}
we denote the longitudinal wavevector mismatch.

The temporal part  can be approximated with the following form \citep{Srivathsan2014}:
\begin{equation}
\psi_{t}(\tau)=\mathcal{N}e^{-\tau/2\tau_{0}}\Theta(\tau),\ \tau=t_s-t_i\label{eq:psi_t}.
\end{equation}
Importantly, the time constant $\tau_{0}$
depends on particular geometrical configuration, i.e. the PM,
exposing the non-trivial connection of the temporal and spatial properties
of the generated state.

The experimental setup details are provided in Fig.~\sref{fig:exp_set}{c}.
The signal, idler and all of the control beams (including Pumps as
well as Write and Seed) are combined in the far field (FF) of the
ensemble and propagate along the $z$-axis.  Since the signal and
idler fields closing the 6WM process have orthogonal polarizations
(respectively: $\sigma^{+}$and $\sigma^{-}$), we separate them using
quarter-wave plate and Wollaston prism. Then after passing the interference
filters (IF), depending on situation, they are either detected on
the Intensified-sCMOS camera \citep{Lipka2018} situated in the far-field
of the ensemble or coupled to single-mode fibers connected to
single photon counting modules (SPCM). To filter out uncorrelated 795 nm photons coming from $5P_{1/2},F=2\rightarrow5S_{1/2},F=2$ decay path,  we employ an additional filtering consisting of glass
cell containing optically pumped to $5S_{1/2},F=2$ state rubidium-87
vapor and buffer gas (nitrogen, 10 Torr). For alignment purposes
we send an additional beam indicated as incoming idler in the Fig.
\sref{fig:exp_set}{c} to seed the signal-idler (SI) generation process
and observe the classically generated signal beam. For the temporal
correlations measurements we first observe the generated signal beam
and maximize its amplitude in the sense of satisfying the PM
condition $\Delta k_{z}=0$. Then we couple this signal beam to the
single-mode mode fiber of the signal SPCM. The particular experimental
configuration i.e. control beam angles and propagation direction was
chosen to provide broad PM spatial spectrum, while the
particular choice of atomic states and polarizations  was motivated by minimization
of loss and noise. Essentially, by selecting $F=3$ manifold as the first
intermediate state $|b\rangle$ we avoid performing read-out of the
SW with the Pump 1 field. Still, the Pump 1 field re-scatters
excitation to other magnetic states of the $F=2$ ground-state manifold
causing effective decoherence of the SW. In the Fig. \sref{fig:exp_set}{d}
we present the time sequence of the experiment including all the steps
described above.

\section{Temporal correlations}
\begin{figure}
	\includegraphics[width=1\columnwidth]{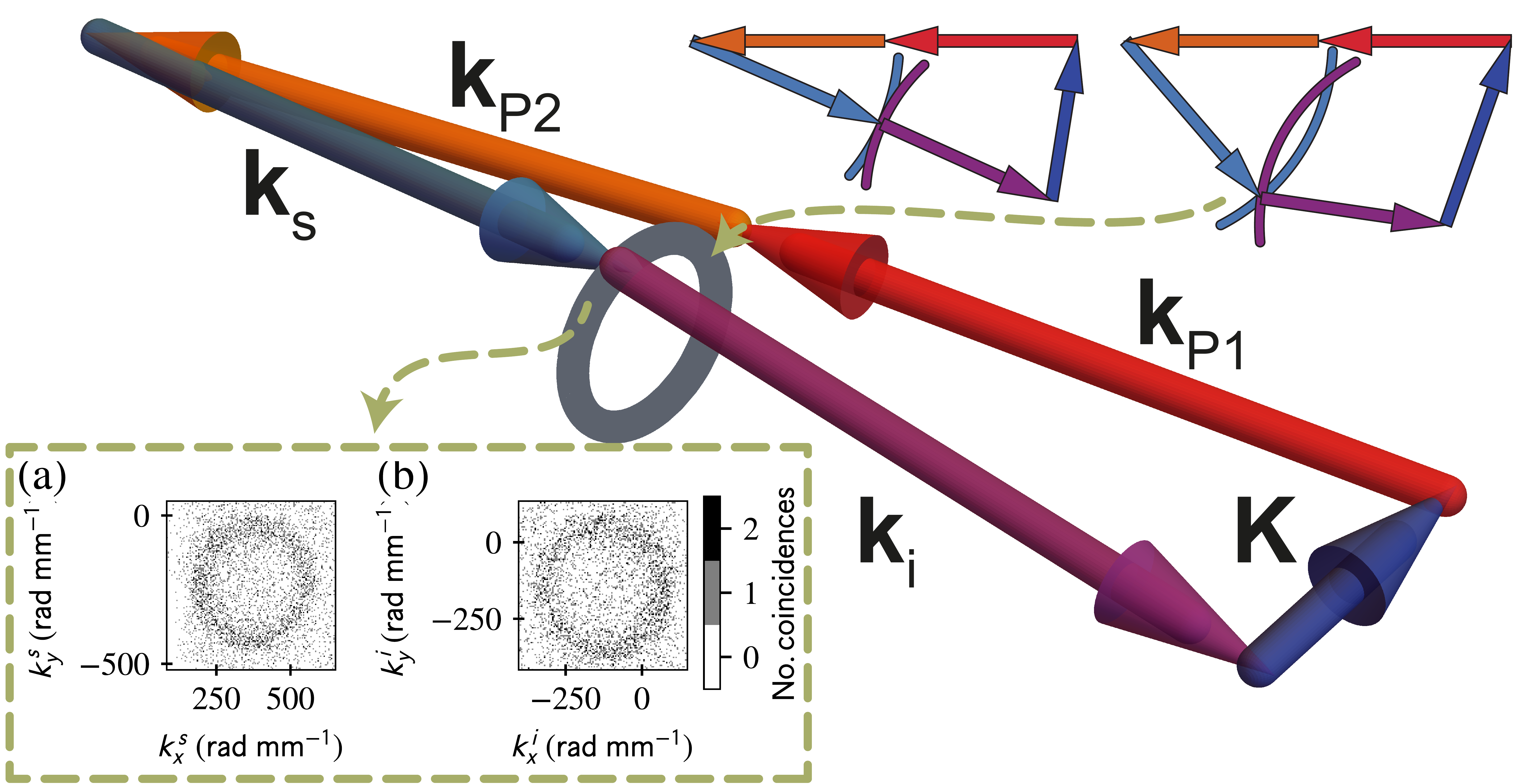}
	
	\caption{Three-dimensional configuration of the wavevectors of all fields interacting
		in the second part of the delayed 6WM scheme.
		The circled by the $\mathbf{k}_{s}$ wavevector region indicates the
		set of SI emission directions $\{\mathbf{k}_{s},\mathbf{k}_{i}$\}
		satisfying the PM condition $\Delta k_{z}=0$. In the insets we plot measured
		signal (a) (idler (b)) photon count distribution conditioned on correlated
		($|\Delta\mathbf{k}_{\perp}|\leq\delta k$) idler (signal) photon
		events.\label{fig:k-conf}}
\end{figure}

We first present the measurement of temporal correlations
in a particular, pre-selected wavevector configuration. Fig. \sref{fig:tcorr}{a}
depicts a histogram of SI correlation as a function of
mutual delay, where we fit the second-order cross correlation as $g_{si}^{(2)}(\tau)=1+\alpha|\psi_{t}(\tau)|^{2}\ast h(\tau)$.
By introducing convolution with $h(\tau)=(\Theta(\tau)-\Theta(\tau-\delta t))/\delta t$
we account for finite time resolution $\delta t=3.85\ \mathrm{ns}$ of the time tagging module. We observe very strong correlation $g_{si}^{(2)}(0)=35.3\pm0.8$, with simultaneously measured auto-correlations $g_{ii}^{(2)}(0)=1.11\pm0.06$
and $g_{ss}^{(2)}(0)=1.13\pm0.16$. These two measurements constitute
a strong violation of the Cauchy-Schwarz inequality \citep{Paul1982}
with:
\begin{equation}
R=\left(g_{si}^{(2)}\right)^{2}/\left(g_{ss}^{(2)}g_{ii}^{(2)}\right)=986\pm153\gg1.
\end{equation}

We measure coincidences during the P1+P2 pulse transferring the ground-state
SW to the exited state $|c\rangle$, that last approx. 300
ns, as well as during a subsequent decay. We observed that if the
excitation pulse is applied continuously, decoherence of the SW $\rho_{hg}$ by pump P1 is strong and the number of coincidences
decreases. Quite similarly, even during the decay of the $|c\rangle$
state we observed that the initial period yields highest signal-to-noise
ratio effectively represented by $g_{si}^{(2)}$. We attribute this
effect to a thermal decay \citep{Zhao2009} of $\esw$. In Fig. \sref{fig:tcorr}{b}
we present additional data showing the temporal evolution of signal-idler
correlation during the two-photon coherence transfer (creation of
$\esw$) and the subsequent decay. 
Indeed, even with negligible $K$ (length of ground-state SW
wavevector), the longitudinal component of $\esw$ is $K_{\esw,z}=2\pi(\frac{1}{\lambda_{P1}}+\frac{1}{\lambda_{P2}})\approx1.6\times10^{4}\ \mathrm{rad/mm}$,
yielding motional characteristic Gaussian decay time of $\tau_{T}^{-1}=|\mathbf{K}_{\esw}|\sqrt{k_{B}T/m_{\mathrm{Rb}}}\approx0.7/\mu\mathrm{s}$. 

We measure the correlation time $\tau_{0}=9.8\pm0.2$ ns which suggest
a strong role of superradiance in the process, as without superradiant
emission enhancement we would expect $\tau_{0}=27.7$ ns, which is a natural lifetime of state $|d\rangle$. Furthermore, thermal decoherence of SW $|h\rangle\langle d|$ and any magnetic decoherence are much slower. With this
we experimentally prove that superradiance is inherently linked to
PM emission \citep{Jen2015}, and can occur in any wave
mixing process. Here, it is demonstrated in 6WM for the first time
to our knowledge.

\section{Wavevector-domain correlations}

\begin{figure*}
	\includegraphics[width=0.99\textwidth]{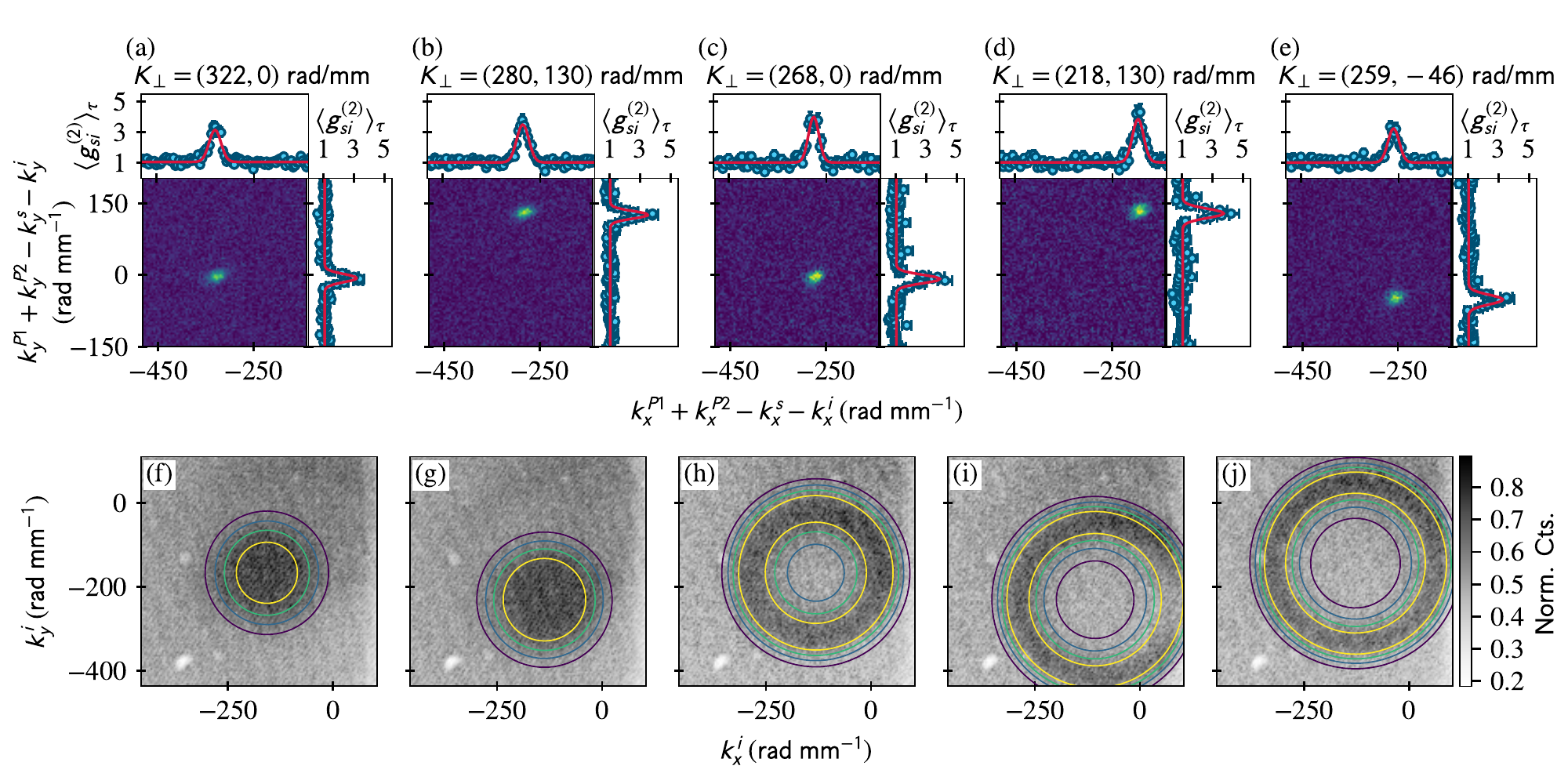}
	
	\caption{Correlation functions and superradiant emission patterns measured
		for a set of the ground-state SW wavevectors. (a-e) Sum of
		momenta coordinates SI second order correlation function
		$\langle g_{si}^{(2)}\rangle_{\tau}$ maps for different ground-state
		SWs. The correlation peak emerges from
		PM condition $\Delta k_{z}=0$ and its position indicates
		the SW transverse components $\mathbf{K}_{\perp}=(K_{x},K_{y})$.
		(f-j) Accumulated idler photon counts for approx. 5M experiment repetitions.
		The contour plots are the theoretical predictions of the PM.
		The data correspond to the same experimental situation column-wise.
		\label{fig:phm-pattern}}
\end{figure*}
In the second experiment we analyze SI correlations in the
wavevector space. Using the I-sCMOS camera we collected 5M frames
for each of five different values of the SW wavevector $\mathbf{K}$.
Here, the $g_{si}^{(2)}(\mathbf{k}_{\perp}=\mathbf{k}_{P1\perp}+\mathbf{k}_{P2\perp}-\mathbf{k}_{s\perp}-\mathbf{k}_{i\perp})$
is calculated in the wavevector-sum coordinates with shifted variables
by the sum of transverse components of pumps wavevectors (see Fig.~\ref{fig:k-conf} for the full wavevector configuration). In this
configuration, we expect that the correlation peak will appear in
the spot $\mathbf{k}_{\perp}=\mathbf{K}_{\perp}$ determined by the initial SW.
This is indeed what we observed in Figs. \sref{fig:phm-pattern}{a-e}.
Note that the correlations are averaged over the entire duration of
the pulse, which we denote by $\langle g_{si}^{(2)}\rangle_{\tau}$.
The averaging, with the pair-generation rate yielding 0.5 signal photon detected per frame on average, leads to significantly lower values for the cross-correlation than in Fig. \sref{fig:tcorr}{a}.
\begin{figure}
	\includegraphics[width=1\columnwidth]{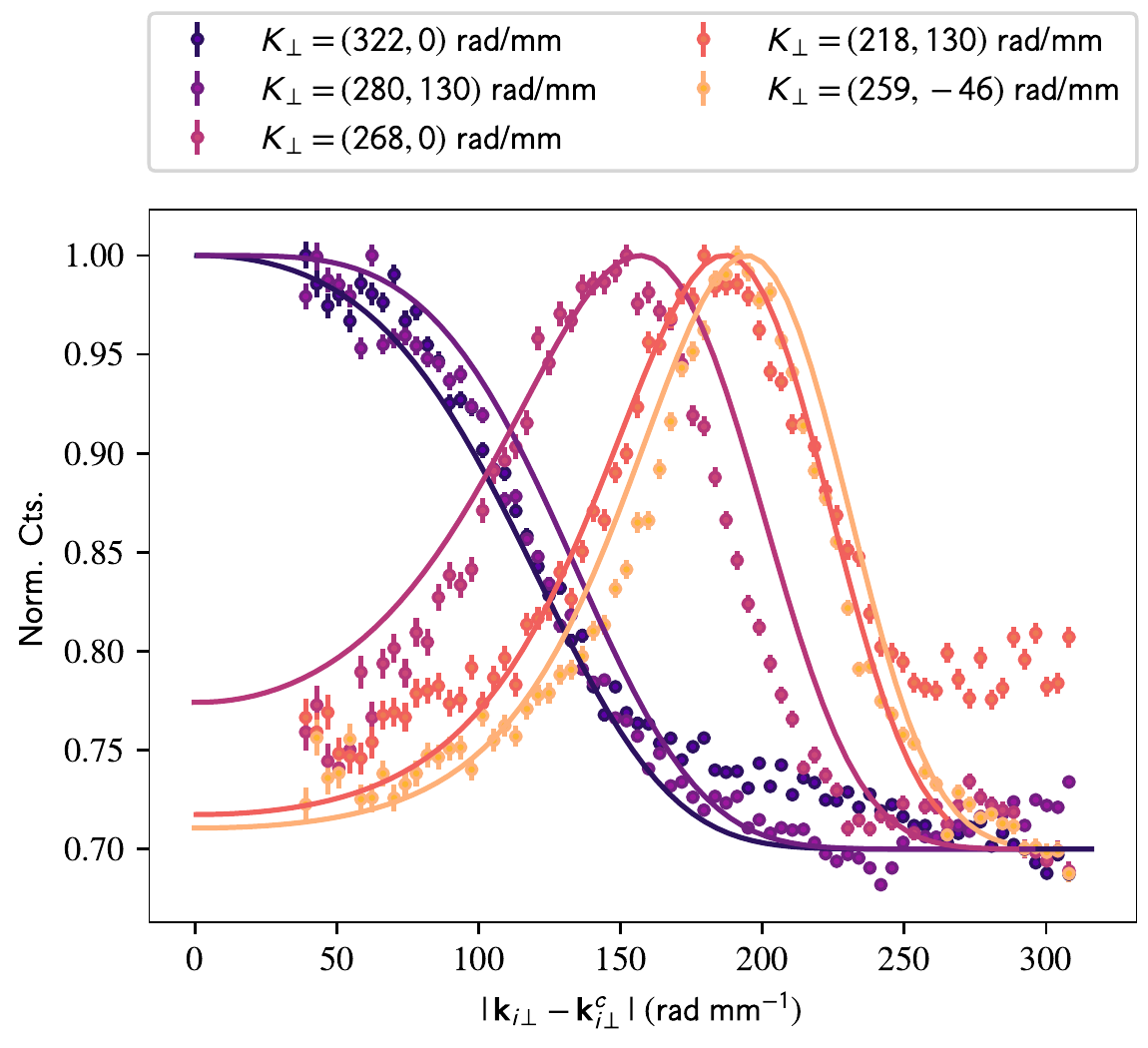}\caption{Average normalized number of idler counts as a function of distance
		$\left|\,\mathbf{k}_{\perp}^{i}-\mathbf{k}_{\perp c}^{i}\,\right|$
		from the center of the circular region $\mathbf{k}_{i\perp}^{c}$
		determined by phase matching cone. The solid lines correspond to the
		theoretical prediction of the phase function described by Eq. \ref{eq:nki}. Discrepancies at the edge are attributed to varying noise background.\label{fig:Average-normalized-number}}
	
\end{figure}
The PM condition renders the SI emission cones
and thus, we expect that correlated signal and idler pairs will appear
in ring-shaped regions on the camera.  To verify that we filter SI
events by choosing only those for which $|\Delta\mathbf{k}_{\perp}|\leq\delta k$, where by introducing
$\delta k=15\ \mathrm{rad\,mm^{-1}}$ we account for a finite spread
of the correlation peak. In insets (a,b) of Fig. \ref{fig:k-conf}
we plot signal and idler accumulated events after the filtering
procedure. We see that the correlated photons mostly come from the
ring-shaped regions.

Finally, we show that the PM patterns are most easily
recovered by looking at the absolute intensity of registered idler
light channel to phase-matched regions by superradiance. The intensity in perfectly correlated regions is increased
approximately two-fold (see Figs. \sref{fig:phm-pattern}{f-j}). This
can be understood as an interplay between coherent and incoherent
emission. The signal photon is emitted in a random direction, leading
to a creation of a random SW $\rho_{dg}$. In
most cases this SW either does not lead to phase-matched emission
(see Fig. \ref{fig:k-conf}) or decoheres rapidly due to atomic motion.
A photon associated with such excitation is thus emitted in a random
direction, in particular also in the PM cone. This cone is quite narrow, and there, the PM
leads to expected coherent emission of idler photon. Thus, starting
with a mean SW number $\bar{n}$ per pixel, we arrive at still
roughly $\eta\bar{n}$ incoherent idler photons per pixel and additional
$\bar{n}_\mathrm{coh}=\eta\bar{n}$ coherent emission into the phase-matched
region. This number, and thus the ratio $(\bar{n}+\bar{n}_\mathrm{coh})/\bar{n}$
may get decreased due to incoherent excitation of SW field $\rho_{hg}$,
incoherent transfer to $\esw$, as well as its motional
decoherence. The actual value which we obtain in the experiment is
$(\bar{n}+\bar{n}_\mathrm{coh})/\bar{n}\approx1.4$.

We compare the expected shape of phase-matched regions described by:
\begin{equation}
\bar{n}_\mathrm{coh}(\mathbf{k}_{i\perp})\propto\int e^{-2\Delta K_{z}(\mathbf{k}_{s\perp},\mathbf{k}_{i\perp})^{2}\sigma_{z}^{2}}\mathrm{d}^{2}\mathbf{k}_{s\perp}\label{eq:nki}
\end{equation}
 with the observed shapes of regions of increased idler counts rate.
Without fitting (since all parameters are determined from correlation
measurement), we observe very good agreement of the shape between
experiment and theoretical prediction.

In Figure \ref{fig:Average-normalized-number} we compare circular
averages of the normalized idler counts for all cases from Fig. \ref{fig:phm-pattern}.
Each point represents an average number of photon counts within a
ring at radius $\left|\,\mathbf{k}_{\perp}^{i}-\mathbf{k}_{\perp c}^{i}\,\right|$ with
unit width, where by $\mathbf{k}_{\perp c}^{i}$ we denote the position
of the center of each phase-matching pattern from Fig. \sref{fig:phm-pattern}{f-j}).
The actual values of $\mathbf{k}_{\perp c}^{i}$ are derived from
Eq. \ref{eq:nki}. With this detailed view we show explicitly that
correlation phase-matching patters, ranging from Gaussian-like shapes
to tight rings can be engineered by controlling the spin-wave wavevector
$\mathbf{K}$. Additionally, one can imagine that even more complex
patterns could be obtained by preparing the initial spin wave in a
superposition in $k$-space or by reshaping the spin-wave spatial structure
using for example ac-Stark shift \citep{Parniak2019,Mazelanik2019,Lipka2019}.

\section{Conclusions and perspectives}

In conclusion we have demonstrated SW based control of superradiance
in the six-wave mixing process. Our experiments paves the way towards
new schemes involving specifically prepared atomic states that lead
to non-standard emission, such as for example sub-radiance \citep{Jen2017a}
as well as interference in the superradiant emission. Interesting
perspectives arise due to the increased range of possibilities in
the wavevector space, compared with low-dimensional interfaces. With
those features, applications such as optical quantum amplifiers \citep{Svidzinsky2013}
or superradiant clocks \citep{Norcia2016} can be extensively studied
and optimized in atom-ensemble systems where SW control is
feasible. Nowadays, extensive studies of superradiance in atomic
arrays or one-dimensional interfaces \citep{Solano2017,Asenjo-Garcia2017,Hood2016,Goban2015,Sorensen2016},
where atoms are strongly coupled to each other, distinctively reveal
associated many-body quantum effects, such as cooperative resonances
and cooperative Lamb shift \citep{Shahmoon2017,Meir2014}. Remarkably
however, even in optically-dense extended free-space atomic ensembles \citep{Srivathsan2013a} as seen in this
work, superradiance plays a crucial role.

Spatial engineering of the demonstrated superradiance extends beyond fundamental aspects. In particular, control of the enhancement of the emission rate in a particular direction helps extend the bandwidth of a photon and potentially match it to a particular receiver. On the other hand, canceling the phase matching allows selective retrieval of stored atomic coherence, and thus temporal-multimodality.

In the more practical context, our experiment shows that photons resonant to excited-to-excited state transitions couple to the ground-state coherence. While preliminary demonstrations for classical light have been made \cite{Parniak2016a,Ding2012}, here we have demonstrated quantum correlations in this scheme. This opens the path to embedding the two functions of a frequency-converted quantum repeater node \cite{Radnaev2010a} in a single atomic ensemble. Practically it is beneficial to use a telecom transition (for rubidium for example 1529 nm and 1475 nm transitions leading to $4\mathrm{D}_{3/2}$ manifold).

Finally, it would be particularly interesting to replace one of the
strong driving fields in the 6WM process by a quantum field. Essentially,
generation of a single SW excitation can be heralded by detecting
a photon scattered in the Raman process. Such configuration directly
leads to generation of a correlated three-photon state. Note that
the process could also be reversed - first, a single SW would
be generated by heralding detection of a photon pair. The resulting
state would be a three-partite entangled state of a photon pair and
a collective atomic excitation. We envisage that both configurations
can be achieved by increasing the two-photon excitation efficiency
- currently limited by decoherence caused by the sole Pump 1 field.
Additionally, it would be also interesting to couple signal field
to the cavity enhancing the emission to the phase-matched region,
preferably in the co-linear SI configuration (Fig. \sref{fig:phm-pattern}{a}).
In this way, the brightness could even surpass the photon down-conversion
probability in three-photon SPDC experiments \citep{Ding2015c,Hubel2010}. Such a tripartite entanglement generation scheme lends itself to two-dimensional quantum repeaters, as proposed by in Ref. \citep{PhysRevA.94.052307}. Our scheme would hold a direct advantage over other three-photon entanglement generation schemes in the quantum repeater context, as one of the photons is directly created as a stored spin wave, akin to the manner of the original DLCZ scheme.
Furthermore, the SW could also be prepared in a deterministic
way using Rydberg blockade \citep{Wei2011,Li2016}, and non-linear Rydberg interactions can play an intricate role in superradiance

\begin{acknowledgments}
We thank K. Banaszek for the generous support and J.~H.~M\"{u}ller for insightful discussion. This work has been
supported by the Polish Ministry of Science ``Diamentowy Grant''
Projects No. DI2013 011943 and DI2016 014846, National Science Centre
(Poland) grants No. 2016/21/B/ST2/02559, No. 2017/25/N/ST2/01163,
and No. 2017/25/N/ST2/00713. The research is supported by the Foundation
for Polish Science, cofinanced by the European Union under the European
Regional Development Fund, as a part of the \textquotedblleft Quantum
Optical Technologies\textquotedblright{} project carried out within
the International Research Agendas programme. M.P. was supported by
the Foundation for Polish Science within the START programme scholarship.
\end{acknowledgments}

\appendix

\section{Experimental setup details}
\label{apx:esud}
We use an elongated ($\sigma_{x}\times\sigma_{y}\times\sigma_{z}=0.3\times0.3\times2.5\,\mathrm{mm}^{3}$)
ensemble of cold ($T\approx22\ \mu\mathrm{K}$) rubidium-87 atoms in
the magnetooptical trap (MOT), described in details in \citep{Parniak2017,Dabrowski2018}.
Before the actual 6WM experiment the atoms are optically pumped to
the $|g\rangle$ state ($5S_{1/2},F=1,m_{F}=1$ , see Fig.1(a)).
To generate the macroscopic ground-state coherence $\rho_{gh}(\mathbf{r})$,
where $|h\rangle$ = $5S_{1/2},F=2,m_{F}=-1$ we use two phase-coherent
optical fields (Write and Seed, both 795 nm). The $\rho_{hg}(\mathbf{r})$
generation process can be understood as a stimulated (seeded with
Seed light) Raman scattering in the $\Lambda$ system driven by the
strong Write light. Here, the $\Lambda$ system is composed of levels
$|g\rangle$, $|h\rangle$ and $|a\rangle$ ($5P_{1/2},F=2,m_{F}=0$).
In such process, both the Seed light and atomic coherence are amplified
at the expense of the Write field. 

\begin{figure}
	\includegraphics[width=1\columnwidth]{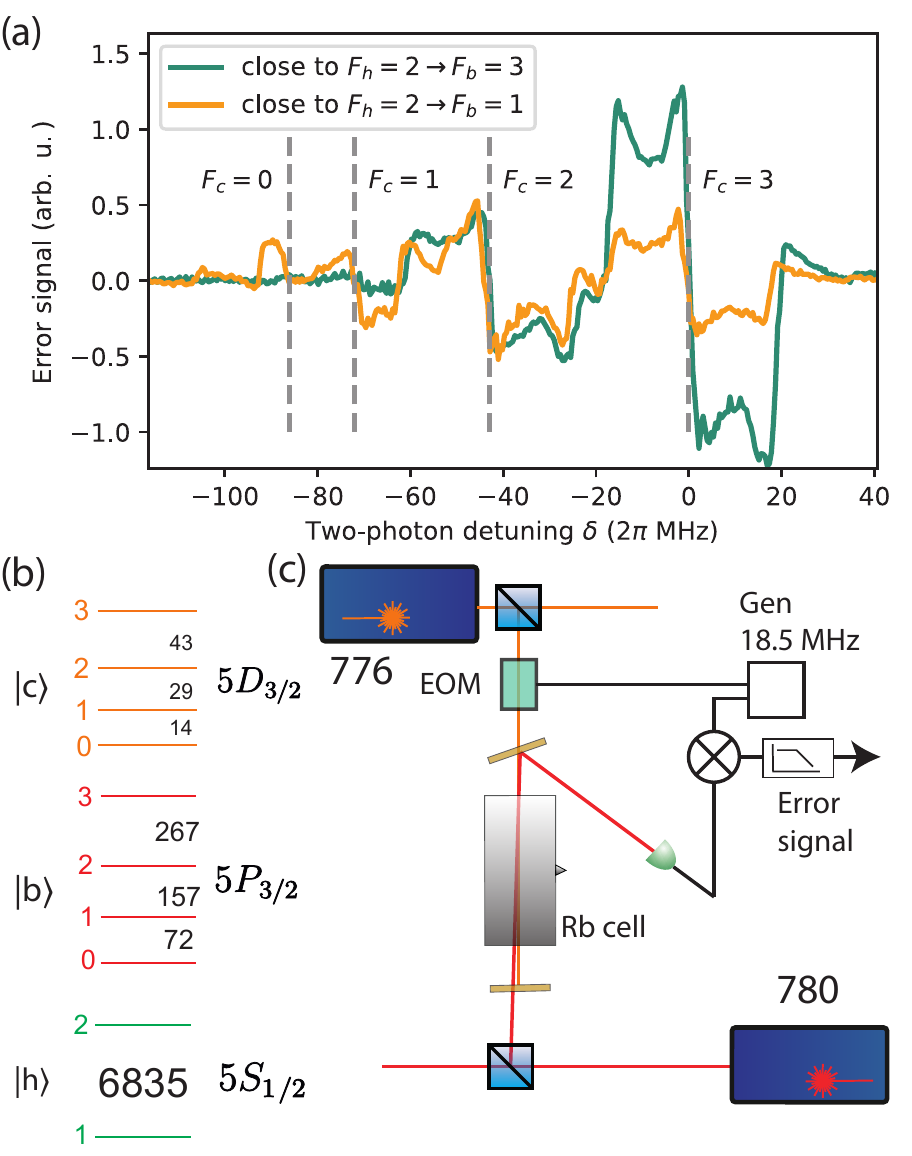}\caption{Modulation transfer spectroscopy setup for locking the 776 nm laser
		to the two-photon resonance. The obtained error signal at two different
		single-photon detunings within the Doppler-broadened line is presented
		in (a). The two-photon detuning is measured from the resonance corresponding
		to $F_{c}=3$. In (b) we show the relevant hyperfine energy levels
		with spacings between them given in MHz. Panel (c) presents the experimental
		setup.\label{fig:Modulation-transfer-spectroscopy}}
\end{figure}

The Seed field is derived from the Write laser using frequency-shifting
setup consisting of electro-optic modulator (EOM) and filtering Fabry-Perot
cavity (see Supplementary Information of Ref. \citep{Parniak2019}). Remaining lasers (excluding
Pump 2) are frequency-stabilized to either cooling-and-trapping laser
(780 nm) or repumping laser (795 nm) using the optical frequency locked
loop described in \citep{Lipka2017}. The Pump 2 laser at 776 nm is
stabilized to the Pump 1 laser using modified version of the two-photon
lock (see Fig. \ref{fig:Modulation-transfer-spectroscopy}) presented
in Ref. \citep{Parniak2016}. The modification employs technique called
modulation transfer spectroscopy, to keep the Pumps 1\&2 in two-photon
resonance between levels $|g\rangle$ and $|c\rangle$.
In this modulation transfer spectroscopy technique \citep{Negnevitsky:13,MartinezdeEscobar:15}, as opposed to the simpler frequency modulation
spectroscopy, we do not modulate the probe light. Instead, we modulate
the coupling (776 nm) light using a phase-only EOM equipped with a
custom resonant circuit, and probe the system with an unmodulated
beam, here at 780 nm (Pump 1 laser). The modulation is thus only transferred
to the probe beam, and detectable in the demodulated signal if the
two beams interact in the vapor cell. The probe beam is then measured
and its RF (18.5 MHz) signal demodulated (mixed) with the pump modulation.
Finally, the error signal is obtained after low-pass filter. A significant
advantage of this scheme is the stability of the error signal offset,
as out of resonance we will simply not have any signal. Furthermore,
we become nearly immune to the residual amplitude modulation of the
EOM, as the amplitude modulation is transduced weakly by the atoms. 

The obtained error signal as we scan the 776 nm laser is presented
in Fig. \sref{fig:Modulation-transfer-spectroscopy}{a}. All four
levels of the highest excited state $|c\rangle$ (Fig. \sref{fig:Modulation-transfer-spectroscopy}{b})
manifold are visible. By changing the single photon detuning within
the Doppler-broadened line corresponding to the $F_{h}=2$ manifold
we observe relative changes of various components that arise due to
selection rules. 

The time domain correlation measurements were performed using Perkin-Elmer
SPCMs ($\sim50 \%$ quantum efficiency) connected to a single-mode
fibers (SMF). The SMF are projecting the output states on a Gaussian
beam modes with waists of radius equal 0.15mm overlapping at the center
of the atomic cloud. The FPGA-based time-tagger which we use provides
3.85 ns time resolution. The spatially resolved detection is performed
using home-built I-sCMOS camera situated in the far field of the ensemble.
The far-field imaging setup was designed using ray-tracing software
(ZEMAX) to maximize the field of view and resolution in $k$-space
(see \citep{Dabrowski2018} for more details). Detailed description
of the I-sCMOS camera itself can be found in Ref. \citep{Lipka2018}.

\section{Spinwave downconversion}
\label{apx:sd}
Treating the atom-light interaction induced by the weak quantum signal
and idler fields as a small perturbation to the free atom density
operator evolution in the presence of $\esw$, we can obtain expressions
for the coherences relevant in the signal idler generation process:
$\rho_{cd}$ and $\rho_{dg}$. Next, we obtain the slowly varying
amplitudes of polarization operators: $\mathcal{P}_{s}=nd_{dc}\rho_{cd}$,
$\mathcal{P}_{i}=nd_{gd}\rho_{dg}$. Then, treating the excited-state
spin wave as one of the interacting fields, we use the relationship:
$\mathcal{P}=\epsilon_{0}\chi E+\epsilon_{0}\chi^{NL}\esw E+...$
to define the linear $\chi$ and nonlinear $\chi^{NL}$ susceptibilities.
Under the assumption of small number of excitations $|\rho_{hg}|\ll1$
and long lifetime of the excited state $|c\rangle$: $\Gamma_{c}\ll\Gamma_{d}$,
we get: 
\begin{gather}
\chi_{s}=0,\\
\chi_{i}=\frac{n}{\epsilon_{0}\hbar}\frac{|d_{gd}|^{2}}{i\Gamma_{d}+2\Delta},\\
\chi_{s}^{NL}=\frac{n}{\epsilon_{0}\hbar}\frac{d_{dc}d_{dg}}{i\Gamma_{d}-2\Delta},\\
\chi_{i}^{NL}=\frac{n}{\epsilon_{0}\hbar}\frac{d_{gd}d_{cd}}{i\Gamma_{d}+2\Delta},
\end{gather}
where $\Delta=\omega_{gd}-\omega_{i}$. As we focus on the signal-idler
generation process only the nonlinear part is of our interest. Especially,
in the interaction picture the effective Hamiltonian for this process
is given by \citep{Wen2007}: 
\begin{equation}
\mathcal{H}_\mathrm{eff}=\epsilon_{0}\int_{V}\mathrm{d}^{3}\mathbf{r}\chi_{i}^{NL}\esw^{+}E_{s}^{-}E_{i}^{-}+h.c.
\end{equation}
where $\esw^{+}=\esw e^{-i\omega_{cg}t}$, and $E_{j}^{-}$ stands
for the quantized fields operators with negative frequency \citep{Wen2007,Wen2007a}
defined \onecolumngrid \noindent as follows: 
\begin{equation}
E_{j}^{-}=\sum_{\mathbf{k}_{j}}E_{j}^{*}a_{\mathbf{k}_{j}}^{\dagger}e^{-i(\mathbf{k}_{j}\cdot\mathbf{r}-\omega_{j}t)}
\end{equation}

\noindent where $E_{j}=i\sqrt{\hbar\omega_{j}/2\epsilon_{0}n_{j}^{2}V_{q}}$
with the quantization volume $V_{q}$ and index of refraction $n_{j}$.
The first-order perturbation theory gives then the state vector:
\begin{equation}
|\Psi\rangle=|0\rangle-\frac{i}{\hbar}\int\mathrm{d}t\mathcal{H}_\mathrm{eff}|0\rangle=|0\rangle+\sum_{\mathbf{k}_{s}}\sum_{\mathbf{k}_{i}}F(\mathbf{k}_{s},\mathbf{k}_{i})a_{\mathbf{k}_{s}}^{\dagger}a_{\mathbf{k}_{i}}^{\dagger}|0\rangle\label{eq:state_w_vac}
\end{equation}
where $F$ is the well known from the SPDC experiments two-photon
spectral function \citep{Rubin1994}, from which we can then obtain
the approximate formulas for temporal (fixed $\{\mathbf{k}_{s},\mathbf{k}_{i}\}$)
and wavevector-domain (time-averaged) wave functions $\psi_{k}(\mathbf{k}_{s\perp},\mathbf{k}_{i\perp})$
and $\psi_{t}(t_{s},t_{i})$ respectively.
\twocolumngrid

\bibliographystyle{apsrev4-1}
\renewcommand*{\bibfont}{\footnotesize}
\footnotesize
\bibliography{bibliografia}

\end{document}